\documentclass[3p, times]{elsarticle}

\usepackage{ecrc}   

\usepackage{ulem}

\volume{00}

\firstpage{1}

\runauth{}

\usepackage{amssymb}

\usepackage[figuresright]{rotating}

\begin{document}

\begin{frontmatter}

\dochead{}

\title{
$K$-series X-ray yield measurement of kaonic hydrogen atoms in a gaseous target
}

\author[label1]{M.~Bazzi}
\author[label2]{G.~Beer}
\author[label3,label4]{G.~Bellotti}
\author[label5,label1]{C.~Berucci}
\author[label1,label6]{A.~M.~Bragadireanu}
\author[label7]{D.~Bosnar}
\author[label5]{M.~Cargnelli}
\author[label1]{C.~Curceanu}
\author[label3,label4]{A.~D.~Butt}
\author[label1]{A.~d'Uffizi}
\author[label3,label4]{C.~Fiorini}
\author[label8]{F.~Ghio}
\author[label1]{C.~Guaraldo}
\author[label9]{R.S.~Hayano}
\author[label1]{M.~Iliescu}
\author[label5]{T.~Ishiwatari}
\author[label10]{M.~Iwasaki} 
\author[label1]{P.~Levi Sandri}
\author[label5]{J.~Marton}
\author[label10]{S.~Okada}
\author[label1,label6]{D.~Pietreanu}
\author[label1,label11]{K.~Piscicchia}
\author[label12]{A.~Romero Vidal}
\author[label1]{E.~Sbardella}
\author[label1]{A.~Scordo}
\author[label1]{H.~Shi\corref{cor1}}
\cortext[cor1]{Corresponding author}
\ead{hexishi@lnf.infn.it} 
\author[label1,label6]{D.L.~Sirghi}
\author[label1,label6]{F.~Sirghi}
\author[label13,label14]{H.~Tatsuno}
\author[label15]{O.~Vazquez Doce}
\author[label5]{E.~Widmann} 
\author[label5]{J.~Zmeskal}

\address[label1]{INFN, Laboratori Nazionali di Frascati, Via E. Fermi 40, I-00044 Frascati(Roma), Italy,                               } 
\address[label2]{Department of Physics and Astronomy, University of Victoria, P.O. Box 1700 STN CNC, Victoria BC V8W 2Y2, Canada,      } 
\address[label3]{Politecnico di Milano, Dipartimento di Elettronica, Informazione e Bioingegneria, Milano, Italy,                      } 
\address[label4]{INFN Sezione di Milano, Via Celoria 16 - 20133 Milano, Italy,                                                         } 
\address[label5]{Stefan-Meyer-Institut f\"{u}r Subatomare Physik, Boltzmanngasse 3, 1090 Wien, Austria,                                } 
\address[label6]{IFIN-HH, Institutul National pentru Fizica si Inginerie Nucleara Horia Hulubbei, Reactorului 30, Magurele, Romania,   } 
\address[label7]{Department of Physics, Facaulty of Science, University of Zagreb, Bijenicka 32, HR-10000 Zagreb, Croatia,             } 
\address[label8]{INFN Sezione di Roma I and Instituto Superiore di Sanita, I-00161 Roma, Italy,                                        } 
\address[label9]{Department of Physics, School of Science, The University of Tokyo, Bunkyo-ku Hongo 7-3-1, Tokyo, Japan,               } 
\address[label10]{RIKEN, Institute of Physical and Chemical Research, 2-1 Hirosawa, Wako, Saitama 251-0198, Japan,                     }  
\address[label11]{Museo Storico della  Fisica e Centro Studi e Ricerche ``Enrico Fermi'', Piazza del Viminale 1-00184Roma, Italy,      }  
\address[label12]{Universidade de Santiago de Compostela, Casas Reais 8, 15782 Santiago de Compostela, Spain,                          }  
\address[label13]{National Institute of Standards and Technology (NIST), Boulder, CO, 80303, USA,                                      }  
\address[label14]{High Energy Accelerator Research Organization (KEK), Tsukuba, 305-0801, Japan,                                       }  
\address[label15]{Excellence Cluster Universe, Technische Universit\"{a}t M\"{u}nchen, Boltzmannstra\ss e 2, D-85748 Garching, Germany.}  

\begin{abstract}
We measured the $K$-series X-rays of the $K^{-}p$ exotic atom in the SIDDHARTA experiment with a gaseous hydrogen target of 1.3 g/l, 
which is about 15 times the $\rho_{\rm STP}$ of hydrogen gas. 
At this density, 
the absolute yields of kaonic X-rays, when a negatively charged kaon stopped inside the target, 
were determined to be 
0.012$^{+0.004}_{-0.003}$ for $K_{\alpha}$ and 0.043$^{+0.012}_{-0.011}$ for all the $K$-series transitions $K_{tot}$. 
These results, together with the KEK E228 experiment results, confirm for the first time 
a target density dependence of the yield predicted by the cascade models, 
and provide valuable information to refine the parameters used in the cascade models for the kaonic atoms. 
}
\end{abstract}


\end{frontmatter}



\section{Introduction}
\label{intro}

The measurements of the $K$-series X-rays 
of kaonic hydrogen $K^{-}p$ and kaonic deuterium $K^{-}d$ atoms 
give uniquely 
the isospin dependent antikaon-nucleon $s$-wave scattering lengths $a_0$ ($I=0$) and $a_1$ ($I=1$). 
Extracted from the energy spectra,
the strong interaction induced energy shifts $\epsilon_{1s}$ and widths $\Gamma_{1s}$ of the ground states of these kaonic atoms
are connected to the $s$-wave scattering lengths via Deser-type formulae \cite{Des54, Mei04}. 
The SIDDHARTA collaboration  
succeeded in giving a precise evaluation of $a_{K^{-}p}$( $=\frac12$[$a_0 + a_1$]) \cite{Baz11}, 
and is preparing to perform the first precision measurement of the $K^{-}d$ X-rays, 
to determine the isospin components $a_0$ and $a_1$ by extracting $a_{K^{-}d}$.

Moreover, 
a new result on the yield of $K^{-}p$ $K$-series X-rays is essential to improve the cascade model calculation of the kaonic hydrogen, 
which is the most poorly understood among all the hydrogen-like exotic $x^{-}p$ atoms, 
where $x^{-}$ represents $\bar{p}, \mu^{-}, \pi^{-}$, or $K^{-}$. 
Up to now only the KEK-PS E228 ($KpX$) experiment obtained $Y_{K_{\alpha}}$ = 0.015 $\pm$ 0.005 and a $K_{\alpha}$ to $K_{tot}$ ratio of 0.27
for a 10 $\rho_{\rm STP}$ hydrogen gas target \cite{Iwa97, Ito98}.
The spectra from three earlier experiments \cite{Dav79, Izy80, Bir83} where a liquid target was used, 
were ambiguous 
and did not allow a reliable subtraction of the number of X-ray events. 
The $K^{-}p$ and $K^{-}d$ $K$-series X-rays are difficult to measure, 
firstly due to their small yields, 
mainly as a result of the Stark mixing of the high-lying atomic states, 
that causes the $K^{-}$ to be absorbed by the nucleus from excited states,  
reducing the rate of the 
transitions to the ground state. 
A second reason is the large natural width of the $1s$ states, 
which makes 
a high signal to background ratio 
hard to achieve.
For $K^{-}p$, 
results of experiments indicate a $1s$ width of $\sim$500 eV \cite{Baz11, Iwa97, Ito98, Bee05};
for $K^{-}d$, 
only theoretical predictions exist, which range from 650 eV to 1000 eV \cite{She12, Dor11, Gal07}.

In this paper, we present the SIDDHARTA experimental result on the kaonic hydrogen $K$-series X-ray yield, 
which together with the $KpX$ result,  
confirms  
a 
density dependence as predicted by multiple cascade calculations \cite{Koi96, Ter97, Jen02, Jen03}. 
The new data will contribute to tuning the parameters including the $2p$ strong-interaction induced width, 
which is used in the latest kaonic hydrogen cascade models as the only free input parameter \cite{Jen02, Jen03, Kal10}.
\section{The SIDDHARTA experiment}
\label{exp}

The SIDDHARTA experiment was performed at the DA$\Phi$NE electron-positron collider, 
where the energies of the beams were tuned to 510 MeV to produce the $\phi$(1020) meson almost at rest in the laboratory frame. 
The 
negatively charged kaons 
from the $\phi\rightarrow K^{+}K^{-}$ decay mode have a low kinetic energy of $\sim$ 16 MeV and a small momentum bite, 
being ideal to be stopped in gas state targets.
The hadronic background at DA$\Phi$NE is substantially lower 
compared to the hadronic beam lines. 
The main component of the DA$\Phi$NE background originating from the beam-beam and beam-gas interactions is composed of minimum ionizing particles (MIPs), 
which can be separated from the slow $K^{-}$ by using the time of flight information. 

\begin{figure}[htbp]
\begin{center}
\leavevmode
   \ifpdf
   \includegraphics[bb =  00 30 250 260, width=9cm]{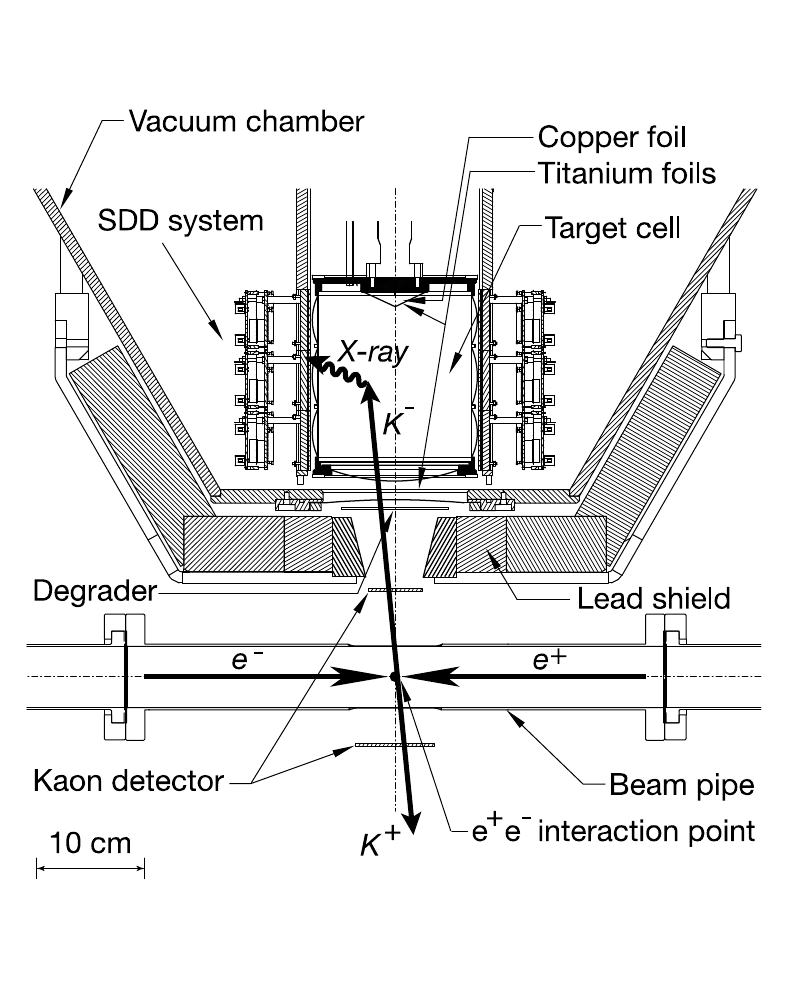}
   \fi
  \caption{
           The schematic view of the SIDDHARTA setup at the DA$\Phi$NE interaction point.
          }
\label{fig:setup}
\end{center}
\end{figure}

As kaon detectors, we used two slices of 1.45 mm thick plastic scintillators placed about 6 cm above and below the interaction point (IP), 
as shown in Fig. \ref{fig:setup}. 
The kaon trigger is defined by the coincidence between the two kaon detectors, 
which gives an identification of the $K^{+}K^{-}$ with an almost back-to-back orientation. 
A clear separation between the kaons and the MIPs was achieved, 
as presented in previous publications of the SIDDHARTA collaboration \cite{Baz11, Baz11KHe3, Baz13Kap}.

Due to the crossing angle of 50 mrad between the two beams, 
the $\phi$ meson has a boost towards the center of the accelerator rings.
To obtain a uniform momentum distribution for the kaons entering the target volume, 
we optimized a mylar degrader to a steps-shape which is thicker towards the center of the accelerator rings \cite{Baz13Kap}. 
After passing through the degrader, the kaons enter the vacuum chamber, 
which contains the target cell and the 144 silicon drift detectors (SDDs), each with an effective area of 1 cm$^2$. 
The stopped $K^{-}$'s form kaonic atoms in highly excited states, 
and the subsequent X-rays from the de-excitations 
are detected by the SDDs with a sub-microsecond timing capability. 
We used the time difference between the SDD events and the kaon trigger to distinguish the kaonic X-rays from the 
beam originating background.
Multiple foils of copper and titanium were placed inside the setup for the energy calibration of the SDDs. 
These foils are irradiated with an X-ray tube from below the setup every few hours in between the measurements with beam. 

For the $K^{-}p$ measurement, 
the cylindrical target cell was filled to 0.13 MPa with hydrogen gas, 
and cooled down to 23 K. 
The DA$\Phi$NE collider delivered a total amount of beam corresponding to 340 pb$^{-1}$ integrated luminosity, 
out of which during the last 106 pb$^{-1}$ of the beam time the experimental setup 
and the DA$\Phi$NE accelerator were operating under stable conditions.
Moreover, the configurations of the degrader, the target, and the kaon detector were finalized and fixed during this last part of beam time, 
after a series of modifications during the first part to optimize the signal to background ratio. 
To extract the yields of the kaonic hydrogen $K$-series transitions, 
we selected the data corresponding to these last 106 pb$^{-1}$ to determine a well-defined systematic error in the comparison to the Monte Carlo simulation.
This data selection criterion is different from, but in no conflict with, 
the previous SIDDHARTA publication \cite{Baz11} in which the complete data set of 340 pb$^{-1}$ was used in order to achieve the best precision in 
the strong interaction shift and width of kaonic hydrogen $1s$ state. 
For the 
data set used in this paper, 
the target density was 1.3 $\pm$ 0.1 g/l, which is about 15 times $\rho_{\rm STP}$ (the standard density of hydrogen gas at 0 $^{\circ}$C and 1 atm).


\section{The absolute yield of $K^{-}p$ X-rays}
\label{sec:yield}
To determine the absolute yield $Y$ of the $K$-series X-rays of the kaonic hydrogen, 
we first determined the detection efficiency of the kaonic X-rays $\epsilon^{\rm EXP}$ from the data, 
by normalizing the number of detected X-ray events ${\rm N_{X\mathchar`-ray}}$ 
with the number of kaon triggers ${\rm N_{ktrg}}$. 
From the Monte Carlo simulation with the Geant4 toolkit \cite{Ago03}, 
we evaluated the $\epsilon^{\rm MC}$,
given by the number of detected X-ray events ${\rm N^{\rm MC}_{X\mathchar`-ray}}$ 
and the number of kaon triggers ${\rm N^{\rm MC}_{ktrg}}$ in the simulation, 
assuming each stopped $K^{-}$ leads to a subsequent kaonic X-ray event. 
The comparison between the two efficiencies gives the absolute yield as : 

 \begin{equation}
 \label{eq:yield}
 {\rm Y } = {\rm \frac{\epsilon^{EXP}}{\epsilon^{MC}}} 
 = {\rm \frac{N_{X\mathchar`-ray}/N_{ktrg}}{N^{MC}_{X\mathchar`-ray}/N^{MC}_{ktrg}} }. 
 \end{equation}

\subsection{Data analysis for ${\rm N}_{\rm ktrg}$ and ${\rm N_{X\mathchar`-ray}}$ }
\label{sec:ana}
In the 106 pb$^{-1}$ of hydrogen data set we selected, 
the number of events identified as kaon pairs at the kaon detectors is ${\rm N_{ktrg}}$ = (1.09 $\pm$ 0.10) $\times$ 10$^7$. 
The MIP events that are miscounted as kaon triggers are less than 1\%. 

\begin{figure}[htbp]
\begin{center}
\leavevmode
   \ifpdf
   \includegraphics[width=9cm]{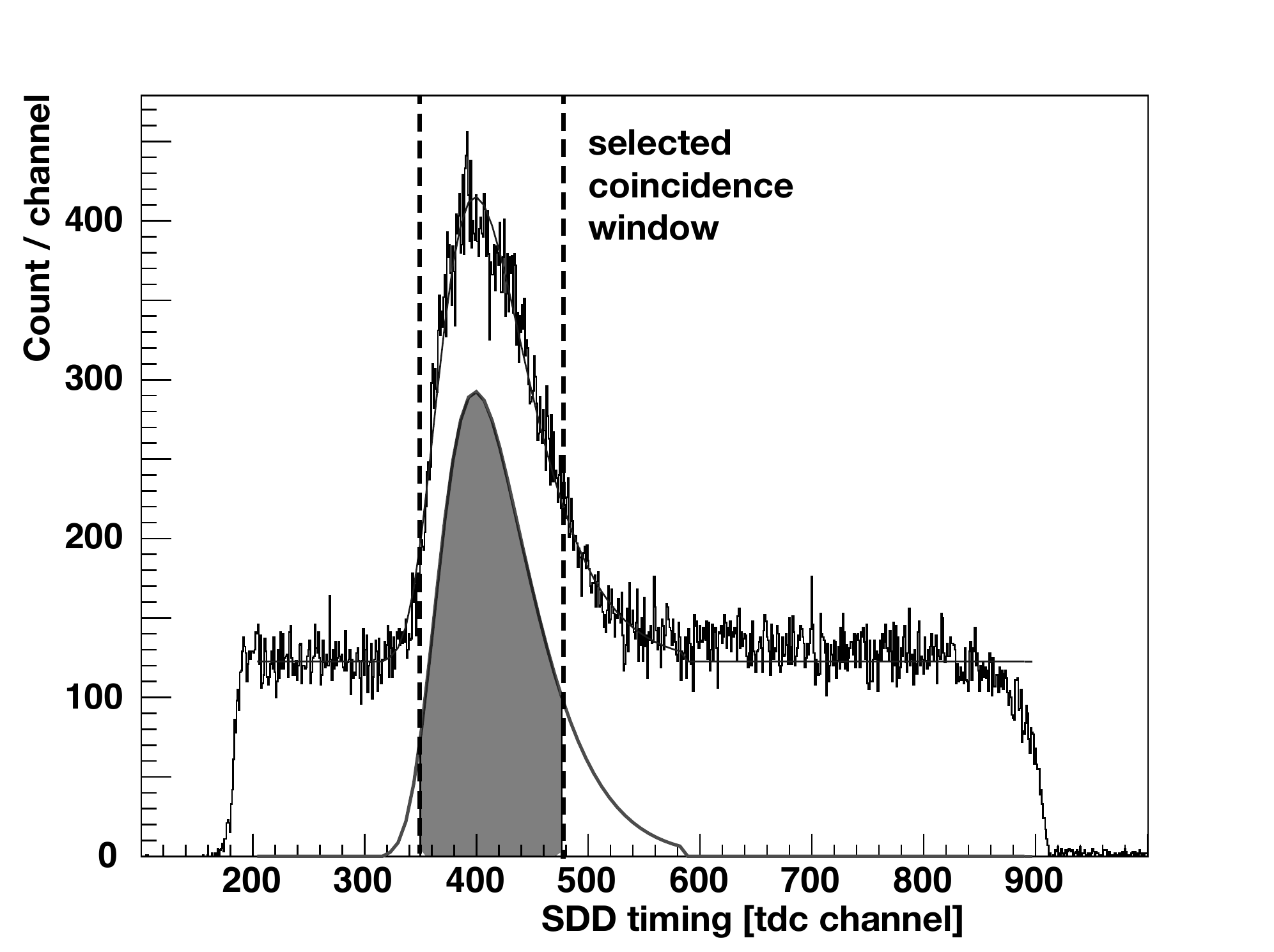}
   \fi
  \caption{
           Timing spectrum of the SDD events. 
           The dashed lines mark the region chosen to get the final kaonic hydrogen energy spectrum.
           Each channel corresponds to 8.3 ns. 
          }
\label{fig:tdc}
\end{center}
\end{figure}

\begin{table}[htbp]
\caption{
        Calculated values of the electromagnetic X-ray energies of the kaonic hydrogen atom, 
         taken into account the vacuum polarization effect, the finite size effect, 
         and the relativistic recoil effect \cite{Kar06, Pil90}.}  
\label{tab:KHX}
\begin{center}
\begin{tabular}[b]{c|c|c}
\hline 
\hline
Transition & X-ray name & Energy (keV) \\ \hline 
$2p\rightarrow 1s$ & $K_{\alpha}$ &  6.481  \\
$3p\rightarrow 1s$ & $K_{\beta}$ &   7.678  \\
$4p\rightarrow 1s$ & $K_{\gamma}$ &  8.096  \\
$5p\rightarrow 1s$ & $K_{\delta}$ &  8.290  \\
$6p\rightarrow 1s$ & $K_{\epsilon}$& 8.395  \\
$7p\rightarrow 1s$ & $K_{\zeta}$ &   8.459  \\
$8p\rightarrow 1s$ & $K_{\eta}$ &    8.501  \\
$\infty\rightarrow 1s$ & $K_{\infty}$    &      8.635 \\  
\hline\hline
\end{tabular}
\end{center}
\end{table}

\begin{figure}[htbp]
\begin{center}
\leavevmode
   \ifpdf
   \includegraphics[bb = 00 00 600 480, width=12cm]{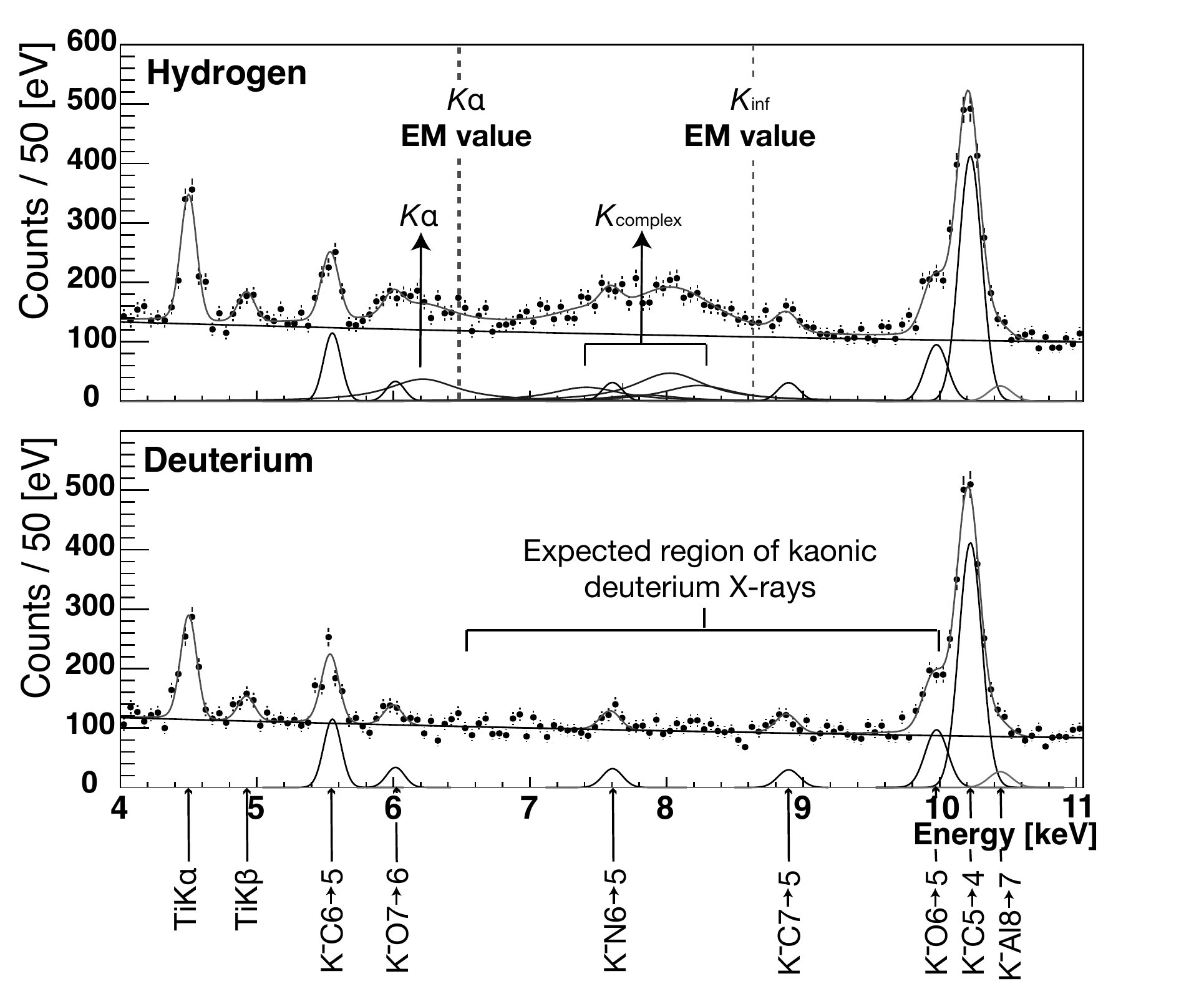}
   \fi
  \caption{Simultaneous fit result of the hydrogen and deuterium targets data.
          }
\label{fig:simfit}
\end{center}
\end{figure}

The energy calibration for each individual SDD was performed using the fluorescence lines of Ti and Cu, 
following the same routine introduced in \cite{Baz11, Baz11KHe3}.
From the data corresponding to 97 SDDs which were operating stably with good energy resolution, 
we selected the events that have a timing coincidence with the kaon trigger as Fig. \ref{fig:tdc} shows,
and obtained the kaonic hydrogen spectrum shown in Fig. \ref{fig:simfit}.
To determine the background of other kaonic X-rays when the $K^{-}$ stopped inside the target gas container made of Kapton polyimide (C$_{22}$H$_{10}$N$_2$O$_5$), 
the energy spectrum from the deuterium target exploratory measurement for 100 pb$^{-1}$ of beam time was included in the analysis, 
as shown in Fig. \ref{fig:simfit}. 

In the analysis we fitted the two spectra simultaneously with a global fit function,  
using the same parameters for the shape of the continuous background over the full energy range, 
and for the relative intensities among the kaonic X-rays from $K^{-}$ stops inside the materials other than the gas volume. 
The parameters for the energy dependent detector resolution are also universal for both spectra. 
Each transition line of the kaonic atom X-rays from $K^{-}$ stops in the Kapton or aluminum in the setup is represented by a Gauss function, 
and the kaonic hydrogen and deuterium lines are represented as the convolution of a Lorentz function and a Gauss function. 
The Lorentzian width corresponds to the width of $1s$ state 
and the Gaussian width represents the detector energy resolution at the mean value of the Lorentz function.
 
To determine the number of kaonic hydrogen X-rays, 
the energy range for the fit was chosen to be from 4 keV to 11 keV, which covers all the kaonic hydrogen $K$-series lines, 
whose expected transition energies considering only the electromagnetic interaction are listed in Table \ref{tab:KHX}.
In Fig. \ref{fig:simfit}, 
the fit result shows a distinguishable $K_{\alpha}$ line 
and a structure called $K_{complex}$ which includes all the transitions to $1s$ from $3d$ and higher energy levels. 
Since the individual lines overlap, 
the relative intensities can not be uniquely assigned. 
To handle the uncertainty among the relative intensities of the $K_{complex}$ lines, 
an iterative fit procedure was applied in the same way as introduced in \cite{Baz11}. 
In the last step of the fit, 
the $1s$ shift and width, and the relative intensities of the higher transitions compared to the intensity of $K^{-}p$ $K_{\alpha}$ are all free parameters, 
with the set values given by the results from the last step in the iteration. 
For kaonic deuterium, the $K_{\alpha}$ and $K_{\beta}$ lines are included in the fit, 
with the $1s$ shift and width fixed to be 500 eV and 1000 eV without losing generality and with their intensities as free parameters. 

As a quality check of the current analysis, the simultaneous fit determines  
the $1s$ state shift and width of kaonic hydrogen of the selected data set to be (only statistical errors): 
$\epsilon_{1s} = -273 \pm 47$(stat) eV, and $\Gamma_{1s} = 567 \pm 103$(stat) eV, 
which are in good agreement with the published results of the SIDDHARTA collaboration  
using 340 pb$^{-1}$ of hydrogen target data \cite{Baz11}: 
$\epsilon_{1s} = -283 \pm 36$(stat) $\pm 6$(syst) eV and   
$\Gamma_{1s} = 541 \pm 89$(stat) $\pm 22$(syst) eV. 

The number of $K_{\alpha}$ events and its statistical error in the $K^{-}p$ spectrum are directly obtained from the results of the fit,
which are listed in Table \ref{tab:kpyield}.
To evaluate the error for the number of $K_{tot}$, however, 
the large statistical errors of other transitions as a result of the correlation among their relative intensities 
must be considered. 
This error assignment was done 
by making use of the covariance matrix associated with the vector of the fit parameters. 
This matrix is given by the Minuit processor MINOS when the minimization of the global Chi-square function is achieved \cite{Jam89}.
More specifically, when we subtracted the parameters that characterize the signal function of the kaonic hydrogen X-rays, 
and integrated over the region of interest to get the number of events, 
the statistical error for the value of the fit function at each energy bin was evaluated using the signal parameter-vector's covariance matrix, 
whose off-diagonal components preserve the correlations among the parameters.
Details of this application in the analysis are discussed in \cite{Shi12}.
To apply this method, all the parameters in the final fit must be free in order to obtain the covariance matrix. 
Thus in the analysis we did not refer to the predictions from any cascade calculations to constrain the relative intensities of the lines in the $K_{complex}$. 
The effectiveness of this method is demonstrated in Table \ref{tab:kpyield}, 
in the sense that the statistical error of the total number of events  
is not affected by the large uncertainties of the individual $K_{complex}$ lines, 
after the correlations are taken into account. 

\begin{table}[tbp]
\caption{
         Results of the intensities for individual and all of the kaonic hydrogen $K$-series transitions from the last step in the iteration 
         of the simultaneous fit. 
        }
\label{tab:kpyield}
\begin{center}
\begin{tabular}[b]{c|c|c}
\hline
\hline
Transition   &  Number of events  &   Statistical error   \\
\hline
$K_{\alpha}$   &  702  &  134 \\
$K_{\beta}$    &  445  &  171 \\
$K_{\gamma}$   &  199  &  521 \\
$K_{\delta}$   &  901  &  930 \\
$K_{\epsilon}$ & 0   &  139  \\
$K_{\zeta}$    & 2   &  136  \\
$K_{\eta}$     & 506 &  357  \\
$K_{tot}$      & 2637   &  395 \\
\hline\hline
\end{tabular}
\end{center}
\end{table}

Another correction to the number of $K_{\alpha}$ and $K_{tot}$ events was done by taking into account that, 
when we applied the events selection based on the SDD timing spectrum in Fig. \ref{fig:tdc}, 
we estimated that the selected region includes about 84\% of the total kaon originating events, 
indicated by the ratio between the grey area and the whole area of the Vavilov function that effectively represented the coincidence events.
As a result, for N$_{\rm ktrg} = (1.09 \pm 0.10) \times 10^7$ kaon triggers in this data set, 
the number of kaonic hydrogen X-ray events we obtained are: 
N$_{K_{\alpha}} = 835 \pm 170$ and N$_{K_{tot}} = 3137 \pm 501$. 
Using the same method with which N$_{K_{tot}}$ is evaluated and applying the same correction for the timing selection, 
the numbers of $K_{complex}$ events is estimated to be N$_{K_{complex}} = 2334 \pm 508$. 

\subsection{Monte Carlo simulation for $\epsilon^{\rm MC}$ }
\label{sec:mc}

A Monte Carlo simulation using the GEANT4 toolkit version 4.9.4 
was done to evaluate the X-ray detection efficiency $\epsilon^{\rm MC}$.
Since in the experiment the positions where the kaons stop are not monitored, 
we combined the simulation for the stopping positions of the kaons and the simulation for the X-ray detection efficiency of SDDs. 
This was realized in the form of a user defined routine in the physics processes for $K^{-}$-at-rest, 
by generating at the exact position where the $K^{-}$ stops 
an isotropically oriented outgoing photon. 
In this way the ratio between ${\rm N_{X\mathchar`-ray}}$ and ${\rm N}_{\rm ktrg}$ 
from the simulation is comparable to the value derived from the experiment. 

In the previous publications of the SIDDHARTA collaboration, 
the profile of the materials that the kaons pass through from DA$\Phi$NE IP until stopping inside the gas volume is listed in detail in \cite{Baz13}, 
and an exhaustive description of the implementation of the physics processes in the simulation was given in \cite{Baz13Kap}.
\begin{figure}[htbp]
\begin{center}
\leavevmode
   \ifpdf
   \includegraphics[width=9cm]{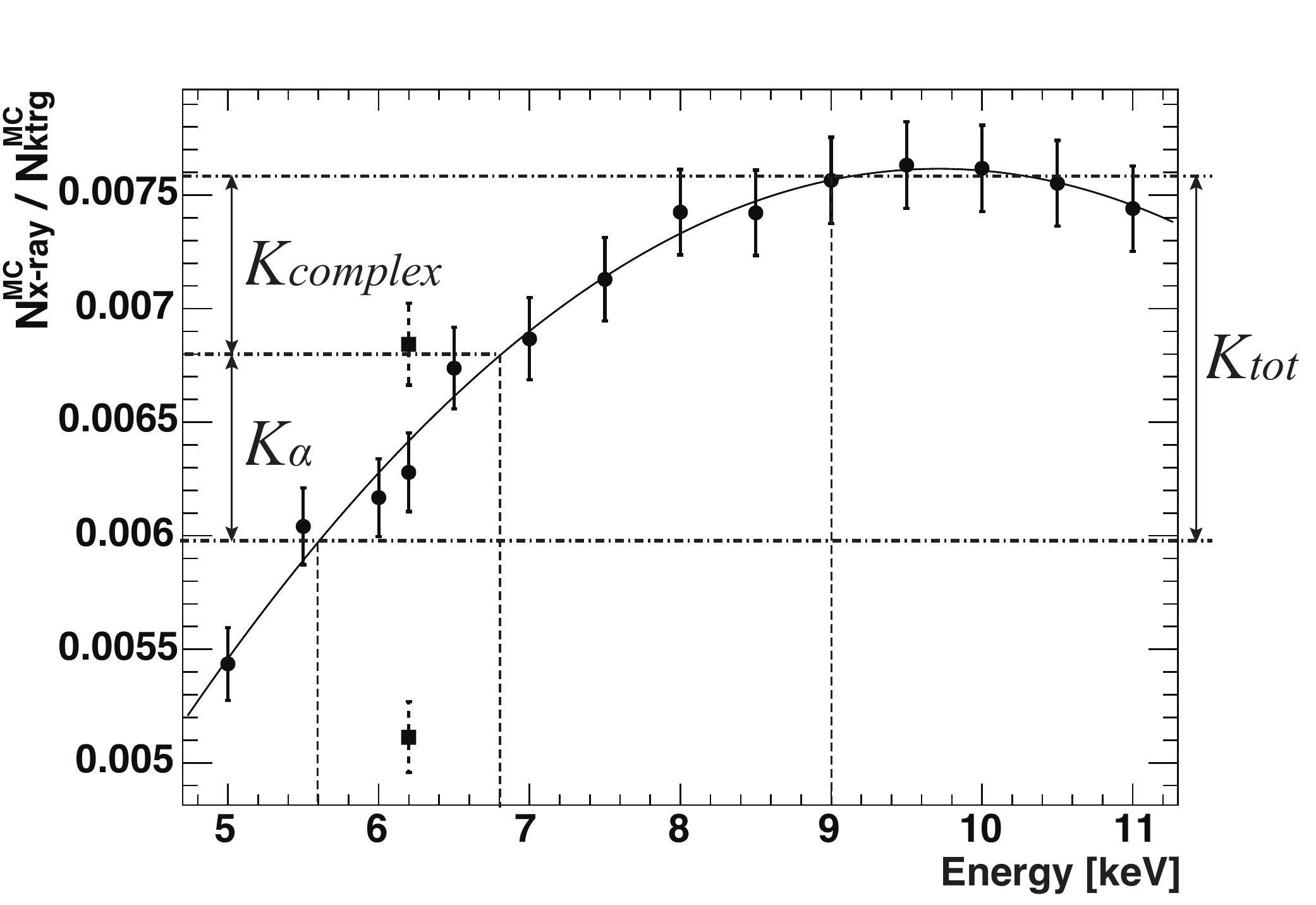}
   \fi
  \caption{The $\epsilon^{MC}$ for different photon energies from MC simulation. 
           The square dots with error bar in dashed line are the results when the thickness of the mylar degrader was varied by $\pm$ 50 $\mu$m.
           The double-arrow segments indicate the range of the systematic errors for $K_{\alpha}$, $K_{complex}$, and $K_{tot}$.
           Details of the plots explained in section \ref{sec:mc}.
          }
\label{fig:mc}
\end{center}
\end{figure}

We initiated the simulation from the generation of the kaon pairs at the DA$\Phi$NE IP, 
with the initial positions smeared 
according to the finite dimension of the crossing region of the beams.
Also we considered the momentum boost due to the beam crossing angle, 
and the angular distribution of the kaons as a result of a spin-one particle decaying into two spin-zero particles. 
The number of events with energy deposit of a few MeV at both kaon detectors 
corresponds to the number of kaon triggers ${\rm N}^{\rm MC}_{\rm ktrg}$.

Consequently, by looking at the energy deposited in the same 97 SDDs as used in the data analysis, 
we counted the number of X-rays that are detected by the SDDs, ${\rm N^{\rm MC}_{X\mathchar`-ray}}$. 
Normalizing with 
${\rm N}^{\rm MC}_{\rm ktrg}$, 
we obtained the detection efficiency from the Monte Carlo simulation $\epsilon^{\rm MC}$.
Over the region of interest from 4 keV to 11 keV, 
we checked the energy dependence of the efficiency, 
based on a series of simulation runs with different energies for the photons generated when the $K^{-}$ stops inside the gas volume.
The photon energy was smeared with a Lorentz function (550 eV width) to reproduce the width of the kaonic hydrogen lines, 
and a Gauss function (60 eV sigma) to reproduce the detector resolution. 
The results are plotted in Fig. \ref{fig:mc}, where the curve is 
a fit to a second order polynomial function. 
For the center value of $K_{\alpha}$ near 6.2 keV, 
the efficiencies when the mylar degrader thickness was varied by $\pm$ 50 $\mu$m (which is the uncertainty of our optimization), are plotted in square dots.

The vertical dashed lines in Fig. \ref{fig:mc} mark the energy boundaries of the $K_{\alpha}$, $K_{complex}$, and $K_{tot}$, 
used to assign their systematic errors of the efficiency by making projections onto the fitted curve. 
For the $K_{\alpha}$ line, which has a distinguishable structure in the energy spectrum, 
the $\epsilon^{\rm MC}$ is given by the simulation run with the photon energy centered at 6.198 keV, 
which is the value taken from the SIDDHARTA result \cite{Baz11}.
The boundaries for $K_{\alpha}$ were chosen as $\pm$ 600 eV from the center energy, 
since they coincide both with the results of experiments for the $1s$ width of about 550 eV, 
and with the energy difference between $K_{\alpha}$ and $K_{\beta}$ as shown in Table \ref{tab:KHX}.
For the $K_{complex}$ and $K_{tot}$, the mean values for the $\epsilon^{\rm MC}$ were chosen, 
without losing generality,
as the average of the vertical arrowed segments marked in Fig. \ref{fig:mc}, 
which are 0.0067 for $K_{tot}$ and 0.0072 for $K_{complex}$, 
with the corresponding systematic errors $\pm$ 13\% and $\pm$ 5\%.
The systematic error of the $K_{tot}$ is noticeably larger, due to a wider energy range.

Other factors that contributed to the final error in the efficiency from the simulation include: 
the uncertainty of the target height ($\pm$ 2\%), 
the uncertainty of the gas density ($\pm$ 2\%) \cite{Shi12}, 
and the statistical error ($\pm$ 2\%).
These errors are added quadratically to the contributions from the degrader thickness uncertainty ($^{+9\%}_{-17\%}$) and the energy dependence of the efficiency. 

From the simulation, the $\epsilon^{\rm MC}$ is estimated to be 0.0063$^{+0.0008}_{-0.0012}$ for $K_{\alpha}$ and 0.0067$^{+0.0011}_{-0.0015}$ for $K_{tot}$.
We also estimated the efficiency for $K_{complex}$ to be 0.0072$^{+0.0008}_{-0.0012}$ for later discussion.

\section{Results and discussions}
By placing 
the numbers 
obtained from sections \ref{sec:ana} and \ref{sec:mc}, 
into Eq. \ref{eq:yield},
we determined the absolute yields for $K_{\alpha}$, $K_{tot}$, and $K_{complex}$ of kaonic hydrogen X-ray as :
\begin{eqnarray}
Y_{K_{\alpha}}  & = & \frac{~835\pm170\ /\ 1.09\pm0.1\times10^7}{0.0063^{+0.0008}_{-0.0012}} = \nonumber 0.012 ^{+0.004} _{-0.003} \nonumber \\
Y_{K_{tot}}     & = & \frac{3137\pm501\ /\ 1.09\pm0.1\times10^7}{0.0067^{+0.0011}_{-0.0015}} = \nonumber 0.043 ^{+0.012} _{-0.011} \nonumber \\
Y_{K_{complex}} & = & \frac{2334\pm508\ /\ 1.09\pm0.1\times10^7}{0.0072^{+0.0008}_{-0.0012}} = \nonumber 0.030 ^{+0.009} _{-0.008}. \nonumber \\
\end{eqnarray}
\begin{figure}[htbp]
\begin{center}
\leavevmode
   \ifpdf
   \includegraphics[width=8cm]{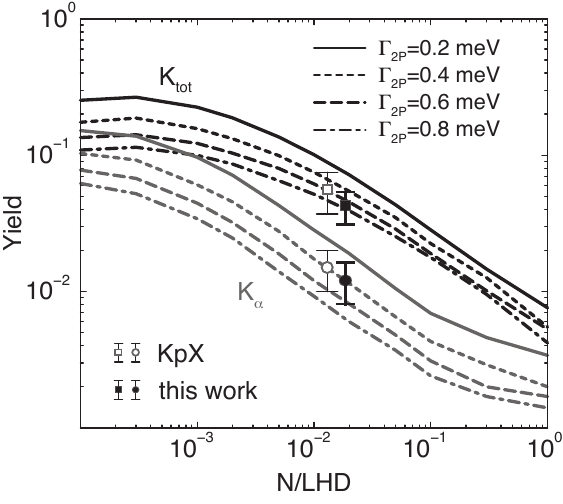}
   \fi
  \caption{The results on the yields of $K^{-}p$ X-rays from two experiments using a gaseous target:
           this work (filled dots) and the $KpX$ experiemnt \cite{Ito98} (hollow dots).
           The theoretical curves are from the cascade calculation by Jensen and others \cite{Jen02, Jen03}, 
           using different $2p$ widths as an input parameter. 
           The horizontal scale is density as a fraction of liquid hydrogen density. 
          }
\label{fig:escm}
\end{center}
\end{figure}
In the extraction of the number of all $K$ transition events from the 
X-ray spectra, 
the uncertainty of the relative intensities of the overlapping $K$-complex transitions 
is evaluated statistically in the analysis, 
without any reference to cascade calculations. 

Our results, together with the absolute yield of $K_{\alpha}$ and the $K_{\alpha}$/$K_{tot}$ ratio given by the $KpX$ experiment, 
confirms the density dependence for the gas density regime predicted by many cascade calculations. 
We take as an example the recent Extended Standard Cascade Model (ESCM) from \cite{Jen02, Jen03} in Fig. \ref{fig:escm}, 
plotted with the results of the $KpX$ and SIDDHARTA experiments.
In this model the broadening effect of the $2p$ state due to the strong interaction $\Gamma^{\rm had}_{2p}$
is the only parameter to be tuned. 
By evaluating the kinematics of the captured kaon at the early stage of the cascade with a classical trajectory Monte Carlo method, 
the Stark effect and the initial kinetic energy of the kaon in the atomic orbit, 
which are introduced as free parameters in the early cascade models \cite{Leo62, Bor80}, 
are quantitatively estimated \cite{Kal10}.
The X-ray yields have been calculated with the ESCM for muonic, pionic, and antiprotonic hydrogen/deuterium atoms, 
and agreements to the experiment data over a wide range of target density were verified by Jensen and others \cite{Jen03}. 
We can expect that a comparison between experiment and cascade calculation results for 
Y($K_{tot}$) /  Y($K_{\alpha}$) and Y($K_{\alpha}$) simultaneously 
may give an estimation of $\Gamma^{\rm had}_{2p}$, as discussed by Jensen \cite{Jen03}.
An improved cascade model will give a more accurate estimate of the $K^{-}d$ X-ray yields, 
which is crucial in planning the beamtime and to optimize the experiment setup of the SIDDHARTA-2 experiment \cite{Sidt2}.

As a final remark, for the discussion about the ratios among the yields, 
we propose to use the absolute yield of the $K_{complex}$ instead of $K_{tot}$ from the experimental point of view, 
because the error for 
Y($K_{complex}$) /  Y($K_{\alpha}$) 
can be quantitatively assigned, thanks to a distinguishable $K_{\alpha}$ line 
from other transition lines.  

\section*{Acknowledgements}
We thank C. Capoccia, G. Corradi, B. Dulach, and D. Tagnani from LNF-INFN; 
and H. Schneider, L. Stohwasser, and D. Stu\"{u}kler from Stefan-Meyer-Institut, 
for their fundamental contribution in designing and building the SIDDHARTA setup. 
We thank as well the DA$\Phi$NE staff for the excellent working conditions and permanent support. 
Part of this work was supported by the European Community-Research Infrastructure 
Integrating Activity ``Study of Strongly Interacting Matter" (HadronPhysics2, 
Grant Agreement No. 227431, and HadronPhysics3 (HP3) Contract No. 283286) 
under the Seventh Framework Programme of EU; 
HadronPhysics I3 FP6 European Community program, Contract No. RII3-CT-2004- 506078; 
Austrian Science Fund (FWF) (P24756-N20); 
Austrian Federal Ministry of Science and Research BMBWK 650962/0001 VI/2/2009; 
Romanian National Authority for Scientific Research, Contract No. 2-CeX 06-11-11/2006; 
the Grant-in-Aid for Specially Promoted Research (20002003), MEXT, Japan;
and the Croatian Science Foundation, under project HRZZ 1680.

\bibliographystyle{elsarticle-num}
\bibliography{references}

\end{document}